\documentclass[twocolumn,showpacs]{revtex4}
\usepackage{graphicx}
\usepackage{dcolumn}
\usepackage{amsmath}
\input epsf.tex
\begin{document}
\title[A drop of hyperfine field at Sn in Fe/Cr/Sn/Cr multilayers]
{A drop of hyperfine field at Sn in Fe/Cr/Sn/Cr multilayers}
\author{A.K. Arzhnikov}
\email{arzhnikov@otf.pti.udm.ru}
\author{L.V. Dobysheva}
\author{D.V. Fedorov}
\affiliation{
Physical-Technical Institute, Ural Branch of Russian Academy of Sciences, \\
Kirov str.~132, Izhevsk 426001, Russia}
\author{V.M. Uzdin}
\affiliation{
St. Petersburg State University, ICAPE,
14 linia V.O. 29, 199178, St. Petersburg, Russia}

\begin{abstract}
The magnetism of the Fe/Cr/Sn/Cr multilayers was studied by
first-principles functional theory. The calculations by LAPW and SKKR
methods showed that two solutions exist in the $Fe_9/Cr_{14}/Sn/Cr_2$
system. One of them is originated from the antiferromagnetic order in
the bulk Cr, and the other is connected with incommensurate spin density
wave (ISDW) in Cr. A characteristic feature of this system is
realization of ISDW in the Cr film thinner than the half length period
of a common ISDW in Cr. In the Cr layers with width lower than a quarter
of the wave, the ISDW cannot be the cause of the sharp drop of the HFF
at Sn. The calculations of the system $Fe_3/Cr_8/Sn/Cr_8$ with and
without allowance for the roughness at the Fe/Cr interface showed that
the roughness leads to a significant decrease in the HFF at Sn.
\end{abstract}
\maketitle
\section{Introduction}

Since the discovery of the antiferromagnetic coupling between Fe layers
\cite{Grunberg} and giant magnetoresistance \cite{Baibich} in Fe/Cr
multilayers a lot of studies have paid a great attention to the Cr
magnetic properties in these systems. Up to now, however, the discussion
on the magnetic order in thin Cr films continues, there is no clarity in 
understanding of the interconnection between the Cr magnetic state and
the magnetic coupling of Fe layers. Recently, the authors of
\cite{Mukhop-PRB} succeeded in obtaining the Fe/Cr multilayer systems
with a Sn monolayer inserted. Multilayer systems
$Fe_{d_1}/Cr_{d_2}/Sn/Cr_{d_3}$ with different widths  $d_1$, $d_2$ and
$d_3$ were investigated by various physical methods including the
M\"ossbauer spectroscopy. One of the most important results was the
established dependence of the hyperfine magnetic field (HFF) at Sn
nuclei on the width of the Cr layer. This dependence is characterized by
a sharp drop of the HFF magnitude at a Cr width less than 3 nm ($d_2+d_3
< 3$ nm) \cite{Almokhtar}. M\"ossbauer experiments cannot give a direct
answer as to the magnetization of Cr atoms, but if the HFF at Sn is
assumed to be proportional to the local magnetic moments of the nearest
Cr atoms, one should inevitably conclude that at a width of the Cr layer
less than 3 nm the Cr magnetic moments in these systems are close to
zero.

As soon as the experimental data had been available, theoretical
first-principles calculations of the electron structure and magnetic
characteristics of Fe/Cr/Sn/Cr were conducted
\cite{Mosida-JMMM,Mukhop-PRB,PRB-2003}. The calculations were performed
by both linear methods \cite{Mosida-JMMM,Mukhop-PRB,PRB-2003} and the
Green function formalism \cite{PRB-2003} with different approximations
of the exchange-correlation potential. As a whole, the results obtained
by different methods and authors agree well with each other and do
confirm the possibility of estimation of the Cr magnetization from
M\"ossbauer experiments at Sn \cite{PRB-2003}. Some  differences between
the results have, though, led to difficulties in explanation of the
peculiarities of the HFF variation and elicited a number of unsolved
questions. 

In the paper presented, we concentrated our attention on the discussion
of two hypotheses advanced earlier in \cite{Mibu-PRL,PRB-2003} for
explanation of the Sn HFF behavior. The first one is connected with the
existence of the high-spin and low-spin states and a transition between
them with a change in the Cr width. Such solutions were obtained for the
$Fe_9/Cr_{14}/Sn/Cr_2$ system in \cite{PRB-2003}. The second hypothesis
is connected with the effect of the Fe/Cr boundary roughness on the Cr
magnetization.

We have performed additional calculations of the periodical systems
$Fe_9/Cr_{14}/Sn/Cr_2$ and $Fe_3/Cr_8/Sn/Cr_8$ (the latter being
calculated with and without account of the roughness of the Fe/Cr
interface). The calculations were conducted by the following methods:
the full-potential linearized augmented plane wave method (FP LAPW)
realized in the program package WIEN2k \cite{WIEN2k}, and a screened
Korringa-Kohn-Rostoker method developed in J\"ulich \cite{Nikos} in the
atomic sphere approximation (ASA SKKR) and in a full-potential scheme
(FP SKKR). The calculations employ the exchange-correlation potential in
the local density approximation (LDA) in the formalism of
Refs.~\cite{Vosko} and \cite{Perdew}. As shown in Ref.~\cite{Cottenier},
if scaled to the moment, the LDA and the generalized gradient
approximation (GGA) yield the same behavior. 

\begin{figure}[!ht]    \epsfxsize=9cm
\centerline{\epsfbox{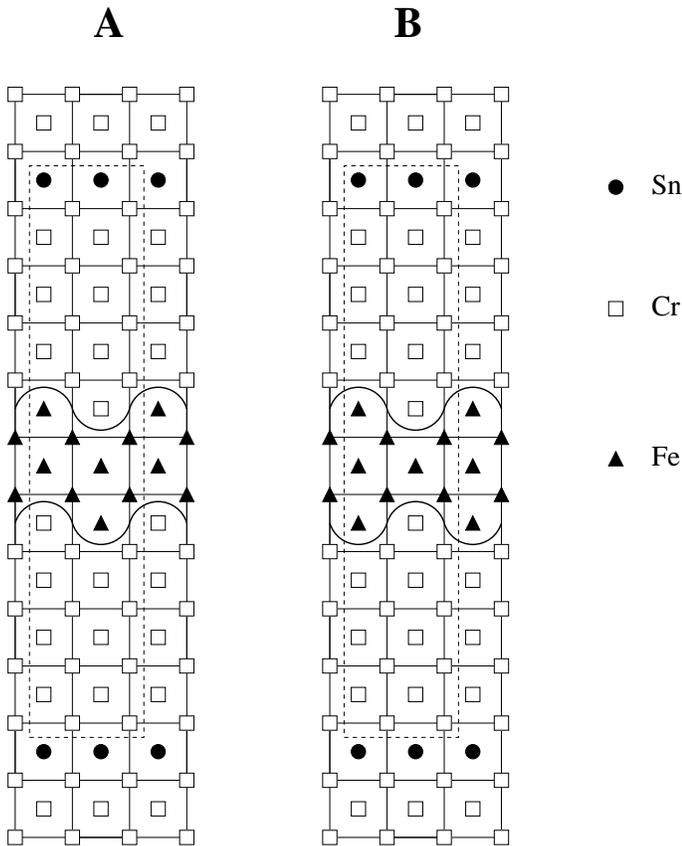}}
\caption{Two types of roughness at the Fe/Cr boundaries (solid lines). 
Dashed lines show the unit cells.}
\hfill{ } \label{roughness}\end{figure}

For our calculations we have used the following models. The multilayer
systems were presented as alternating layers in the (001) plane of a bcc
lattice with two types of roughness (Fig.~\ref{roughness}, A and B) and
without it.
The interlayer distances Fe-Fe, Fe-Cr, Cr-Cr and Cr-Sn were taken equal
to 0.144 nm, which corresponds to the half lattice parameter of the bulk
Cr (0.288 nm). Notice that the lattice relaxation and small variations
of the interlayer distances do not change the qualitative behavior of
the results of Ref.~\cite{PRB-2003}.

\section*{Results}

As stated earlier, the calculation of the $Fe_9/Cr_{14}/Sn/Cr_2$ system
by the SKKR method gave two solutions \cite{PRB-2003} that we called
"high-spin" (HS) and "low-spin" (LS) in accordance with the magnetic
moment magnitudes (the difference in energy between the two solutions
did not exceed 1 mRy per unit cell). In connection with these
solutions it is appropriate to raise here some questions. First: why the
LS solution was not found in Ref.~\cite{PRB-2003} by the LAPW method?
Second: do there exist two solutions in other systems and do they give a
possibility to describe the peculiarities of the Sn HFF behavior as a
transition from the HS to the LS state with a decrease of the Cr width?
And the third question concerns the nature of the LS state (the
character and peculiarities of the behavior of the Cr magnetic moments
in the HS solution were considered in detail and explained in
\cite{PRB-2003}). 

Two solutions were also found in an isolated 23-layer Cr film by other
authors \cite{Bihlmayer}. One of them was classified as a spin density
wave and was connected with incommensurate spin density wave (ISDW) of
bulk Cr. The ISDW has been studied from sixties, and now reliable
experimental and theoretical results are available for both bulk Cr, and
some multilayer films with Cr, the majority of which are reflected in
reviews \cite{Fishman,Zabel} in a rather full measure. We did not meet
in literature with ISDW in Cr films thinner than 20-22 layers. This
seems natural as the ISDW half period  exceeds 20 layers. A hypothesis has,
nevertheless, arisen that the nature of the low-spin state obtained in
\cite{PRB-2003} is connected with ISDW. For the existence of the ISDW
solution, the Fermi-surface topology and the accuracy of its description
are important. Assuming that in the FP LAPW calculation of the work
\cite{PRB-2003} the accuracy of the Fermi-surface description was
unsufficient, we have increased the number of basis wave functions
limiting it by the parameter $R_{MT} K_{max} = 7.5$ (in \cite{PRB-2003}
it was $R_{MT} K_{max} = 7.0$). The expansion of the wave functions over
the angular momentum was done up to $l_{max}=10$, and the Fourier
expansion of the electron charge density was conducted up to
$G_{max}=20$. We should note that in the modification of the method used
here, namely, the augmented plane wave plus local orbitals (APW+l.o.),
the number of plane waves is usually limited by $R_{MT}K_{max} = 7.0$,
as the convergence is achieved at lower number of basis functions than
in the common realization of the LAPW method \cite{Madsen}. We have
increased also the number of k-points in the irreducible part of the
Brillouin zone up to 42, which may be crucial for the description of the
Fermi surface topology.  The calculation was conducted in a
scalar-relativistic approximation for the valence electrons and it was
fully relativistic for the core electrons with an energy boundary at -7
Ry separating these two groups. 

\begin{figure}[!ht]    \epsfxsize=9cm
\centerline{\epsfbox{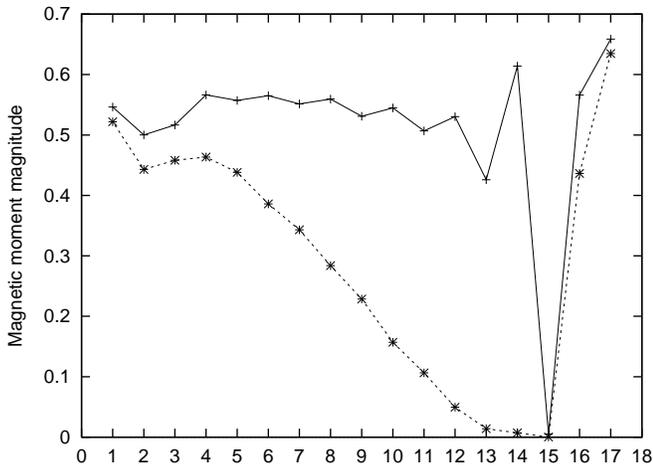}}
\caption{The high- and low-spin solutions found by the FP LAPW method 
for the $Fe_9/Cr_{14}/Sn/Cr_2$ system}
\hfill{ } \label{fig2}\end{figure}

The results of calculation of the $Fe_9/Cr_{14}/Sn/Cr_2$ system are
given in Table~\ref{Table1} and Fig.~\ref{fig2}. 
\begin{table}
\caption{
The magnetic moments ($\mu_B$) in $Fe_9/Cr_{14}/Sn/Cr_2$.
\label{Table1}}
\begin{ruledtabular}
\begin{tabular}{c|c|c|c}
Atom & layer &  HS      &     LS \\
  \colrule
Fe &   &   2.260    &    2.277 \\
Fe &   &   2.317    &    2.326 \\
Fe &   &   2.343    &    2.346 \\
Fe &   &   2.382    &    2.386 \\
Fe &   &   2.101    &    2.087 \\
Cr &17 &  -0.659    &   -0.635 \\
Cr &16 &   0.566    &    0.437 \\
Sn &15 &   0.005    &    0.000 \\
Cr &14 &   0.614    &   -0.007 \\
Cr &13 &  -0.426    &   -0.014 \\
Cr &12 &   0.530    &    0.050 \\
Cr &11 &  -0.507    &   -0.106 \\
Cr &10 &   0.545    &    0.157 \\
Cr & 9 &  -0.531    &   -0.229 \\
Cr & 8 &   0.560    &    0.284 \\
Cr & 7 &  -0.551    &   -0.343 \\
Cr & 6 &   0.565    &    0.386 \\
Cr & 5 &  -0.557    &   -0.438 \\
Cr & 4 &   0.566    &    0.463 \\
Cr & 3 &  -0.517    &   -0.458 \\
Cr & 2 &   0.500    &    0.443 \\
Cr & 1 &  -0.546    &   -0.522 \\
Fe &   &   2.034    &    2.021 \\
Fe &   &   2.429    &    2.429 \\
Fe &   &   2.330    &    2.343 \\
Fe &   &   2.340    &    2.352 \\
           \end{tabular}
	     \end{ruledtabular}
		 \end{table}
The calculations were conducted with a charge convergence criterion 
$10^{-4} \div 10^{-5}$ (this is a squared charge-density difference
between the last two iterations integrated over the space). As seen in
Fig.~\ref{fig2}, in the FP LAPW method at the chosen parameters we have
succeeded in obtaining two solutions, one of them being the LS solution.
The HS solution is practically the same as in \cite{PRB-2003} by the FP
LAPW method. The difference in energy between the HS and LS solutions is
1.4 mRy per cell. We should mention a very slow convergence of the LS
solution, which prevented us from obtaining the accuracy over spin
density of $10^{-5}$ as for the HS solution (the accuracy of $10^{-4}$
for LS was obtained). The LS solution obtained here has a peculiarity
absent in the solution obtained by the SKKR method in
Ref.~\cite{PRB-2003}, it is a phase slip between the 13-th and 14-th Cr
atoms (if counted from the Fe layer, see Table~\ref{Table1}). The "phase
slip" usually denotes a violation of the alternation of the magnetic
moment directions characteristic for the antiferromagnetic ordering, and
reveals itself in the fact that two neighboring Cr layers have equally
directed magnetic moments. It is a common knowledge that the phase slip
is one of the main features of the ISDW state \cite{Zabel}. To obtain an
additional argument for identification of the LS solution as an ISDW
state, we have recalculated the $Fe_9/Cr_{14}/Sn/Cr_2$ system by the
SKKR method with an accuracy higher than in \cite{PRB-2003}. 

For the SKKR calculation a screening potential with a barrier height of
4 Ry is used and the structure constants include the coupling to the
six neighboring spheres. A cutoff of the angular momentum at
$l_{max} = 3$ is used for the Green function, that implies a cuttof of
the charge-density (in the case of ASA and FP) and potential (in the
case of FP) components at $2 l_{max}= 6$. The shape of the Voronoi cells
for the FP calculations is described by the proper shape functions
expanded up to $4 l_{max} = 12$.

The energy integration of the Green function for the determination of
the electron density in the self-consistency procedure is performed by
means of the Gaussian quadrature with 33 points on a semicircle (with 5
Matsubara poles). For the $k$ integration we use four kinds of meshes with 
182, 45, 28, and 15 $k$ points in the two-dimensional irreducible part 
of the Brillouin zone.

\begin{figure}[!ht]    \epsfxsize=9cm
\centerline{\epsfbox{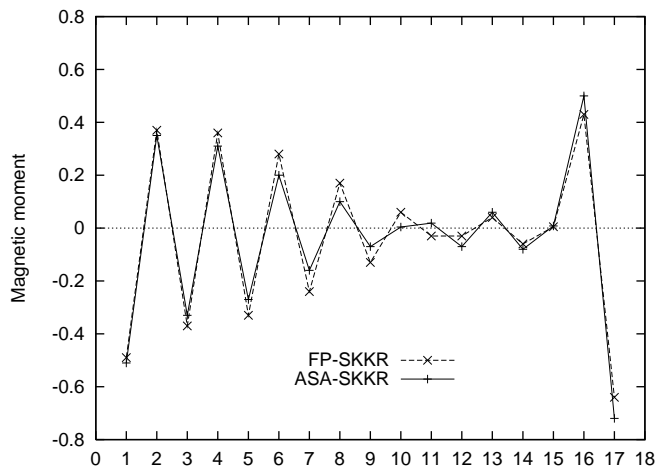}}
\caption{
The ISDW solutions found by the ASA and FP SKKR methods}
\hfill{ } \label{fig3}\end{figure}

\begin{figure}[!ht]    \epsfxsize=9cm
\centerline{\epsfbox{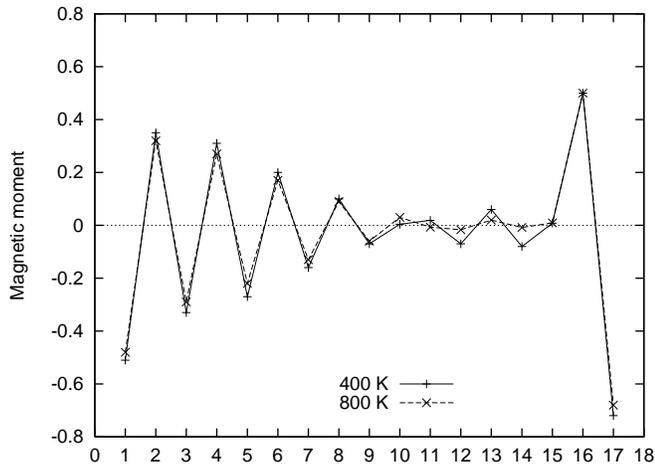}}
\caption{The ISDW solutions found by the ASA SKKR method at temperatures 
400 and 800 K}
\hfill{ } \label{temp}\end{figure}

The accuracy of the convergence over the potential was  $10^{-6} -
10^{-7}$ Ry, which is an order higher than in \cite{PRB-2003}. The
increase of the number of points in these four kinds of $k$-mesh up to
9800, 3150,  975 and 342, respectively, as well as the use of the
double-dense energy mesh do not practically change the results. The
calculation results are given in Fig.~\ref{fig3} and ~\ref{temp}.

From Fig.~\ref{fig3} one may see that SKKR solutions also have a phase
slip characteristic for the ISDW (between the 10-th and 11-th sites in
ASA SKKR and between the 11-th and 12-th sites in FP SKKR). Note that
here we did not succeed in obtaining the second, HS, solution by the
SKKR method with an accuracy better than in \cite{PRB-2003}.

Fig.~\ref{temp} shows the calculation results at two temperatures (400
and 800 K). We should note that to describe the magnetic state
variations with temperature (including the phase transitions), the
Heisenberg excitations should be allowed for. Our calculations take into
account, with the help of the temperature Fermi distribution function,
only Stoner-type excitations, which become effective at temperatures as
high as $5000 - 10000$ K \cite{Moriya}. This allows us to suppose that,
at 400 and 800 K used, the Stoner excitations weakly affect the magnetic
moment magnitude, and to consider the temperature as a mean to change
the details of the Fermi surface that determine the position of the
phase slip. Indeed, from simple considerations, we may suppose that the
ISDW period is determined by the position of the chemical potential
which depends on the temperature (see, for example,
\cite{Ashcroft}) 
$$\mu=\epsilon_F-\pi^2 k_BTn'(\epsilon_F)/6 n(\epsilon_F) +        
o(k_BT/\epsilon_F)$$  
\noindent where $\epsilon_F$ is the Fermi energy, $k_B$ is the Boltzmann
constant, $n(\epsilon_F)$ and $n'(\epsilon_F)$ are the electron density
of states at the Fermi level and its derivative (the value of
$n'(\epsilon_F)$ is negative in our system). With increasing
temperature, the chemical potential grows, decreasing thus the distance
in the k-space between the electron and hole surfaces which determines
the ISDW period. That is why the phase slip moves from the 10-11-th
sites to the 11-12 sites with the temperature change from 400 to 800 K
(see Fig.~\ref{temp}).

The above results leave no doubt that the nature of the LS solution in
the $Fe_9/Cr_{14}/Sn/Cr_2$ system is connected with the ISDW state in
Cr. The realization of this state in our system has, however, some
features absent in the ISDW in other systems \cite{Fishman,Zabel}. The
Sn layer present in the system gives the possibility to form a state
with a quarter of the ISDW, which was not observed in theoretical
calculations earlier. Furthemore, the role of the Sn layer in the ISDW
formation is here different from that found in \cite{Mibu-PRL2002} in
the Cr/Sn multilayers, where Sn layers are pinning centers for the
spin-density wave maxima.

Proceeding from the nature of the LS state one may assert that its
realization in the multilayer systems with a Cr layer width less than a
quarter of the ISDW (11-12 Cr layers) is impossible, so the hypothesis
about the existence of the two solutions (HS and ISDW) in thinner Cr
layers and the transition between them with a decrease of the Cr width
is untenable. This is confirmed also by calculations of the systems with
thin Cr layers \cite{PRB-2003}, where no low-spin solution was found.

So, the second explanation of the experimental dependence of the HFF at
Sn connected with the effect of the Fe/Cr boundary roughness becomes
more preferable. To check this hypothesis we have considered two models
of the roughness of the Fe/Cr boundary shown in Fig.~\ref{roughness}.
The calculations for the film $Fe_3/Cr_8/Sn/Cr_8$ were conducted by the
FP LAPW method, the results are given in Table~\ref{Table2}.
\begin{table}
\caption{
The HFF at Sn (T) and magnetic moments ($\mu_B$) in $Fe_3/Cr_8/Sn/Cr_8$
\label{Table2}}
\begin{ruledtabular}
\begin{tabular}{c|c|c|c|c}
 Atom  &layer&ideal&A roughness (see Fig.~\ref{roughness})&B roughness\\
  \colrule            
 Sn HFF&     & -16.1  &         -5.0          &-4.7 / -5.0\footnote[1]
 {.../... denotes values for two nonequivalent atoms in the layer} \\
  \colrule            
  Sn   &  11 &  0.01  &         0.004         &  0.004  /   0.004\footnotemark[1]\\ 
  Cr   &   1 &  0.73  &         0.23          &        0.23        \\    
  Cr   &   2 & -0.48  &-0.15 / -0.15\footnotemark[1]&-0.15 / -0.15\footnotemark[1]\\
  Cr   &   3 &  0.60  &         0.18          &        0.18        \\ 
  Cr   &   4 & -0.57  &-0.18 / -0.18\footnotemark[1]&-0.16 / -0.18\footnotemark[1]\\
  Cr   &   5 &  0.60  &         0.19          &        0.17        \\
  Cr   &   6 & -0.55  &-0.18 / -0.24\footnotemark[1]&-0.15 / -0.22\footnotemark[1]\\
  Cr   &   7 &  0.54  &         0.23          &        0.19        \\
Cr / Fe&   8 & -0.60  &-0.53 /  1.92\footnotemark[1]&-0.50 /  1.89\footnotemark[1]\\
  Fe   &   9 &  1.97  &         2.25          &        2.25        \\
  Fe   &  10 &  2.46  &         2.43          & 2.37 / 2.47\footnotemark[1]\\
           \end{tabular}
	     \end{ruledtabular}
		 \end{table}
One may see from Table~\ref{Table2} that the roughness in both model
cases leads to a decrease in magnetic moments of the Cr atoms nearest to
the Sn layer. Tis results in a significant decrease in HFF at Sn from
16.1 T (for the system with ideal interfaces) to 4.7 - 5.0 T (for the
systems with roughness). This does not explain in full measure the
experimental sharp drop of the HFF. To obtain it, one should perform the
calculation of the systems with higher width and different roughnesses. 

\section*{Summary}

The first-principles FP LAPW calculation of the multilayer model system
$Fe_9/Cr_{14}/Sn/Cr_2$ gives two solutions. One of them, with a lower
magnitude of the Cr magnetic moments and a phase slip, was identified as
an ISDW. This identification was confirmed by the SKKR calculations. A
characteristic feature of the ISDW state in our case is the fact that the
Sn layer promotes the realization of the ISDW state in the films with
Cr width less than the half period of ISDW.

In connection with the above statements, one may assert that the band
calculations of the periodical multilayer systems without the interface
roughness cannot explain the experimentally observed sharp drop of the
HFF at Sn with a decrease of the Cr layer width.

The presence of roughness in the Fe/Cr interfaces leads to an HFF drop
in thin films. This is confirmed by the first-principles FP LAPW
calculations conducted for the model $Fe_3/Cr_8/Sn/Cr_8$ with ideal Fe/Cr
interfaces and with two types of roughness. The calculations showed that
there is a threefold drop in HFF in the systems with roughness as
compared to the system with ideal Fe/Cr boundaries. However for
explanation of the experimental dependence it is necessary to calculate
a number of systems with more than 20-23 Cr layers and with allowance
for the different roughness.

\section*{Acknowledgments}

One of us (D.V.F.) would like to thank Dr. Nikos Papanikolaou and
Professor Ingrid Mertig for providing a version of the screened KKR code
and valuable advice. 
This work was partially supported by INTAS (grants 01-0386 and
03-51-4778), and RFBR (grants 03-02-16139 and 04-02-16024 )

\end{document}